\documentclass[%
 reprint,
superscriptaddress,
 amsmath,amssymb,
 aps,twocolumn
]{revtex4-2}
\usepackage{float}
\usepackage{graphicx}% Include figure files
\usepackage{subfigure}
\usepackage{dcolumn}% Align table columns on decimal point
\usepackage{bm}% bold math
\usepackage{hyperref}% add hypertext capabilities
\usepackage{natbib}%
\usepackage[compatibility=false]{caption}

\begin{document}

\preprint{APS/123-QED}

 \title{In-plane Exciton Polaritons vs Plasmon Polaritons: \\Nonlocal corrections, confinement and loss\\}
%\title{Interband vs Intraband collective excitations in atomically thin layers:\\Nonlocal effects, confinement and loss \\}%
%\title{Nonlocal Corrections in Atomically Thin Layers: A Comparison of Interband and Intraband Collective Excitations}
% Force line breaks with \\

\author{Yonatan Gershuni}
\affiliation{School of Electrical Engineering, Faculty of Engineering, Tel Aviv University, Tel Aviv 6997801, Israel}
\affiliation{School of Physics and Astronomy, Tel Aviv University, Tel Aviv, 6997801, Israel}
\affiliation{Center for Light-Matter Interaction, Tel Aviv University, Tel Aviv 6997801, Israel}
\affiliation{QuanTAU, Quantum Science and Technology Center, Tel Aviv University, Tel Aviv 6997801, Israel}
\author{Itai Epstein}
\email{itaieps@tauex.tau.ac.il}
\affiliation{School of Electrical Engineering, Faculty of Engineering, Tel Aviv University, Tel Aviv 6997801, Israel}
\affiliation{Center for Light-Matter Interaction, Tel Aviv University, Tel Aviv 6997801, Israel}
\affiliation{QuanTAU, Quantum Science and Technology Center, Tel Aviv University, Tel Aviv 6997801, Israel}


\begin{abstract}
Polaritons are quasi-particles describing the coupling between a photon and a material excitation, which can carry large momentum and confine electromagnetic fields to small dimensions, enabling strong light-matter interactions. In the visible (VIS) to near-infrared (NIR) spectral ranges, the intraband response of metals gives rise to surface-plasmon-polaritons (SPPs), which have practically governed polaritonic response and its utilization in nanophotonics. Recently, the concept of interband-based VIS/NIR in-plane exciton polaritons has been introduced in two-dimensional materials, such as transition-metal-dichalcogenides (TMDs), thus providing an excitonic alternative to plasmonic systems. Here, we compare the properties of such in-plane exciton polaritons supported by monolayer TMDs to the equivalent configuration of SPPs supported by thin metallic films, known as the short-range-SPPs (SRSPPs). Taking into account both excitonic and plasmonic nonlocal corrections, which play a major role in large momentum modes, we find that in-plane exciton polaritons provide confinement factors that are an order of magnitude larger than those of SRSPPs, and with six times lower propagation losses. In addition, we show that unlike SPPs, in-plane exciton polaritons are coupled to the TMD's valley degree of freedom, leading to directional propagation that depends on the exciton's valley. These properties make in-plane exciton polaritons promising candidates for VIS/NIR nanophotonics and strong light-matter interaction.

%\begin{description}
%\end{description}
\end{abstract}

\maketitle
%\tableofcontents

\section{Introduction}
In-plane propagating polaritonic waves describe the coupling between electromagnetic fields and matter excitations that can be electronic, phononic or excitonic, depending on the material properties. These polaritonic waves can be supported and waveguided by interfaces between two different materials, by thin layers, or in the bulk. Examples of such polaritons include surface-plasmon-polaritons (SPPs) at metal/dielectric interfaces \cite{Zayats2005,MAIER2007THEBOOK}, graphene-plasmons (GPs) supported by monolayer graphene \cite{Koppens2011,Grigorenko2012,Iranzo2018,Epstein2020_Far}, and hyperbolic phonon polaritons (HPhPs) in bulk anisotropic materials \cite{Caldwell2014,Dai2014,Ma2018}. SPPs emerge as collective oscillations of conduction electrons in metals, coupled with the electromagnetic oscillations of the photons in the dielectric material. In a similar manner, in semi-metallic graphene, the electromagnetic field couples with the collective oscillation of the charge carriers in the graphene, giving rise to GPs, and HPhPs arise due to the coupling of photons with optical phonons and lead to a hyperbolic dispersion for propagating modes in the bulk. 

The prerequisites for the existence of such polaritonic waves is a negative real part of the permittivity for one of the materials at the interface and a positive for the other. For hyperbolic bulk waves, an anisotropic permittivity is required between the in- and out- of-plane directions. Polaritonic waves of this nature are renowned for their ability to confine light below the diffraction limit, supporting large momenta. These unique properties of in-plane polaritons have lead to numerous exciting discoveries and innovative applications in imaging beyond the diffraction limit, light modulators, sensing, and on-chip photonics \cite{Fang2005,Pacifici2007All-opticalDots,Anker2008BiosensingNanosensors,Fu2012,WangX2013,Nikolajsen2004,Reinhard2011}. \par

Electronic response in material excitations may be divided into two categories. Intraband, describing the electronic behavior of metallic materials via their Drude response, and interband transitions depicting electronic transitions from one band to another, such as exciton formation. SPPs, are a prime example of intraband polaritons, influenced by the properties of both the metal and dielectric. Characterized by large in-plane momentum, SPPs dominate the visible (VIS) to near-infrared (NIR) spectral ranges, which can be traced back to the inherent Drude response of metals \cite{Hass_PhysThinFilms,MAIER2007THEBOOK}.\par

Considering a more complex configuration of a dielectric/thin metallic slab/dielectric, two SPPs can be supported by the two metal/dielectric interfaces. If the metal slab thickness is comparable to the out-of-plane decay length of each SPP, it leads to an overlap and hybridization of the two SPP modes, and the formation of two new modes which are distinguishable by even and odd parities of the in-plane electric field. The even mode, is characterized by a longer propagation length than that of the single SPP but with smaller confinement, while the odd mode, named short-range-surface-plasmon-polaritons (SRSPP) is characterized by larger confinement and higher losses \cite{Berini2000,Berini2001,Jacob2008,Sarid1981-SRSPP,MAIER2007THEBOOK,Dionne2005}. These large losses dramatically hinder SPP-based technologies \cite{Barnes2003SurfaceOptics,MAIER2007THEBOOK}, together with difficulties in fabricating large-area, high-quality, ultra-thin metal sheets or dielectric spacers \cite{Sun2018TuningSystems,Woolf2009TheWaveguides,Jacob2008}, with the current state-of-the-art being 3 nm ultra-thin single-crystalline Au sheets, supporting mid-IR SPPs \cite{Maniyara2019TunableFilms}.

Recently, interband excitonic-based in-plane polaritons have drawn great scientific attention, achieving both  plasmonic and hyperbolic responses \cite{Epstein2020HighlySemiconductors,Eini2022Valley-polarizedFrequencies,Sternbach2021,Wang_Basov_2021}. More specifically, monolayer TMDs encapsulated by hexagonal-boron-nitride (hBN) at cryogenic temperatures exhibit negative real part of their permittivity, mirroring the conditions in metals necessary for the propagation of surface waves \cite{Epstein2020HighlySemiconductors}. In achieving a plasmonic-like response, they have been shown to foster 2D (in-plane) exciton polaritons (2DEPs) with large momentum and relatively modest losses \cite{Epstein2020HighlySemiconductors}. The fundamental difference being that while metals involve electromagnetic field coupling to electrons, TMDs facilitate a robust coupling with excitons. However, a systematic comparative study between 2DEPs and their plasmonic counterparts, the SRSPPs, remained unexplored until now. \par 

Here, we study and compare the fundamental polaritonic properties of 2DEPs and SRSPPs in terms of their confinement factors and propagation losses. For comparison, we study the 2DEPs supported by monolayer TMDs encapsulated by hBN and a benchmark case of SRSPPs supported by a 3 nm single-crystalline Au slab. We incorporate nonlocal corrections for both the excitonic and the plasmonic systems, arising from the pronounced momentum of these polaritonic modes, and find that 2DEPs exhibit confinement factors up to an order of magnitude greater and demonstrate losses reduced sixfold. Our investigation allows an in-depth understanding of the nonlocal 2DEP dispersion relation, pivotal for studying and realizing such polaritons in a lab environment. Moreover, we show that unlike SRSPPs, the studied 2DEPs are coupled to the monolayer semiconductor’s valley degree of freedom, enabling access to directional, valley-controlled polaritonic waves.
\begin{figure}[ht]
    \centering
    \includegraphics[width=1\linewidth]{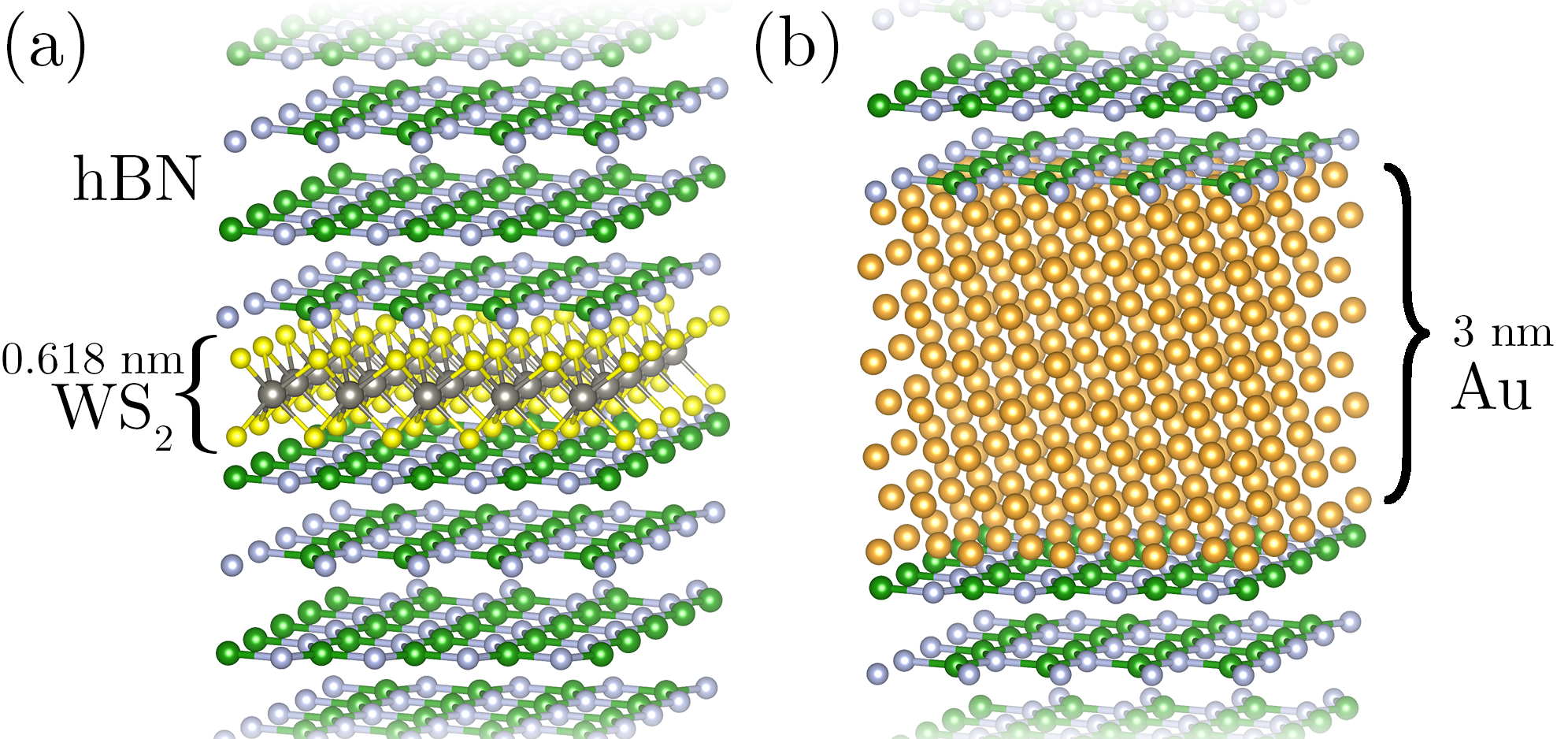}% Here is how to import EPS art
    \caption{Illustrated systems under investigation. (a) Monolayer WS\textsubscript{2}, (b) 3 nm Au, both encapsulated with 
 semi infinite layers of hBN. Illustrated using VESTA.}
    \label{fig:Geometry}
\end{figure} 

%\pagebreak

\section{Nonlocal response of monolayer semiconductors}
In its more general form, the susceptibility of a material is a function of both frequency and momentum. While it is always present in this general description, in most cases the contribution of the momentum to the susceptibility can be neglected, as the momentum of the modes is relatively small, i.e. residing around the light line of the material. Thus, the susceptibility in these cases can be described as dependent on frequency only \cite{MAIER2007THEBOOK}. However, in cases where large momentum modes are supported, such as polaritons with a dispersion relation reaching momentum values that are much larger than the light line, the momentum dependence cannot be neglected, thus necessitating nonlocal corrections to the susceptibility's local response \cite{Lundeberg2017TuningPlasmonics,Goncalves2021QuantumPlasmons,Iranzo2018,Epstein2020_Far}.\par
%\cite{Rajabali2021PolaritonicInteraction,Sciesiek2020Long-distanceMicrocavities,Itskos2007EfficientHeterostructures}. These are the length scale discussion cites\par
The susceptibility of a monolayer TMD including its excitonic response can be locally modeled as a summation over Lorentzian resonances. This is a direct consequence of each exciton behaving as an individual Lorentzian resonance, which takes the form:\cite{Glazov2014ExcitonDichalcogenides,Yu2010FundamentalsProperties,Epstein2020HighlySemiconductors} 
\begin{align}\label{local_SC_susceptibility}
    \chi_\textsubscript{TMD}\left(\omega\right)&=-\sum_j\frac{c\gamma_{r,0_j}}{\omega_{0_j} d_0\left(\omega-\omega_{0_j}+i\left(\gamma_{nr_j}/2+\gamma_{d_j}\right)\right)},
\end{align}
where $c$ is the speed of light, $\omega_0$ is the exciton resonance energy, $d_0$ is the thickness of the TMD and $\gamma_{r0}$, $\gamma_{nr}$, $\gamma_d$ are the radiative, nonradiative, and pure dephasing decay rates, respectively \cite{Epstein_Unity_2020,Eini2022Valley-polarizedFrequencies}. Nonlocality stems from effects an electric field at point $\mathbf{r'}$ has on electric displacement vector $\mathbf{D}(\mathbf{r})$ for $\mathbf{r}\ne\mathbf{r'}$ in the medium \cite{Agarwal1974}. The nonlocal corrections to Eq. \ref{local_SC_susceptibility} can be understood in the classical analogy of allowing a coupling term between neighbouring oscillators \cite{Hopfield1963TheoreticalCrystals}. From a quantum mechanical consideration, spatial dispersion is achieved by allowing a kinetic energy term, $E_k=\frac{\hbar^2 k^2}{2M}$, to the excitonic energy in the Hamiltonian, where $k$ is the exciton momentum and $M$ is the exciton effective mass \cite{Yu2010FundamentalsProperties}. This leads to a momentum dependent exciton eigenenergy in the form of 
\begin{align}\label{correction_to_local_model_exciton}
    E_{ex}\approx\omega_0\rightarrow\omega_0+\frac{\hbar^2 k^2}{2M},
\end{align}
and Eq. \ref{local_SC_susceptibility} becomes
\begin{align}\label{eq: non_local_SC_susceptibility}
    \begin{split}
        \chi_\textsubscript{TMD}\left(\omega,k\right)&=\\
        =-\sum_j&\frac{c\gamma_{r,0_j}}{\omega_{0_j} d_0\left(\omega-\left(\omega_{0_j}+\frac{\hbar^2}{2M}k^2\right)+i\left(\frac{\gamma_{nr}}{2}+\gamma_{d_j}\right)\right)}.
    \end{split}
\end{align}
Focusing on the excitonic resonance $\omega_{0j'}$, the contribution of other higher energy resonances in the summation to the susceptibility can be lumped together into a single constant background term, $\chi_{BG}$, and written as (see supplementary information)
\begin{align}\begin{split}\label{non_local_SC_susceptibility_single}
    \chi_\textsubscript{TMD}\left(\omega,k\right)&=\\
    =\chi_{BG}-&\frac{c\gamma_{r,0}}{\omega_0 d_0\left(\omega-\left(\omega_0+\frac{\hbar^2}{2M}k^2\right)+i\left(\frac{\gamma_{nr}}{2}+\gamma_{d}\right)\right)}.
\end{split}\end{align}
This formulation will be pivotal in deriving the nonlocal dispersion for a 2DEP sustained by a WS\textsubscript{2} semiconductor.

%\pagebreak

\section{Nonlocal response of metal surfaces}
An idealized metal can be described by free-moving electrons within the bulk when subjected to an electromagnetic field, without mutual interactions. In this scenario, the electron density at the metal's surface can be arbitrarily high. In a real metal, electrons close to the surface experience collective interactions resulting in a nonlocal behaviour, such as Coulomb interaction and the quantum Pauli exclusion principle, which prevent electrons from congregating indefinitely at the surface. Instead, they get distributed over a certain volume adjacent to the surface \cite{Ciraci2012ProbingEnhancement}. In ultra-thin metal films and for large momentum modes, for which the wavelength is much greater than the electron mean free path in the material, these nonlocal effects play a decisive role. To accurately account for these interactions, especially at the surface, we employ a hydrodynamic model of light-matter interactions. This model introduces a pressure-like term to the free electron model, accounting for electron interactions. It prevents infinitely high electron densities and places a bound on confinement. The dynamics under this hydrodynamic approach can be expressed as \cite{Moreau2013ImpactAntennas,Ciraci2013HydrodynamicProblem}
\begin{align}\label{Hydromodel}
-\beta^2\mathbf{\nabla}\left(\mathbf{\nabla} \cdot \mathbf{P}\right)+\mathbf{\ddot{P}}+\gamma\mathbf{\dot{P}} &=\varepsilon_0\omega_p^2\mathbf{E},
\end{align}
where $\mathbf{E}$ is the electric field, $\mathbf{P}$ is the polarization vector, $\omega_p=\sqrt{\frac{ne^2}{\varepsilon_0 m_e}}$ is the plasma frequency, $\gamma$ a damping factor and $\beta\propto v_F$, is a nonlocal parameter from the Thomas-Fermi theory of metals \cite{Ashcroft1976SolidPhysics} where $v_F$ is the Fermi velocity. For gold nanostructures $\beta\approx 1.27\cdot 10^6 \text{ m/s}$ has been found to be in good agreement with experimental data \cite{Ciraci2012ProbingEnhancement,Moreau2013ImpactAntennas} accounting for nonlocal effects. Incorporating this with Maxwell's equations leads to a corrected Ampere's law of the form of 
\begin{align}
    \begin{split}\label{eq: AmpereEquation}
        \mathbf{\nabla}\times\mathbf{H}&=-i\omega\mathbf{D(\mathbf{r})}\\
        &=-i\omega\epsilon_0\left(1-\frac{\omega_p^2}{\omega^2+i\gamma\omega}\right)(\mathbf{E}-\\ &\frac{\beta^2}{\omega_p^2-\omega^2-i\gamma\omega}\mathbf{\nabla}\left(\mathbf{\nabla}\cdot\mathbf{E}\right)).
    \end{split}
\end{align} 
The two sets of possible solutions for these equations correspond to transverse and longitudinal modes satisfying $\mathbf{\nabla}\cdot\mathbf{E}=0$ and $\mathbf{\nabla}\times\mathbf{E}=0$ respectively. The transverse mode is the familiar solution to Maxwell's equation which disregards nonlocality and satisfies $\epsilon\frac{\omega^2}{c^2}=k^2$ while the latter solution however, relates to electron oscillations inside the bulk which yields the following dispersion:
\begin{align}\label{eq: metal_longitudinal_wavevector}
    %\frac{\beta^2}{\omega_p^2-\omega^2-i\gamma\omega}\nabla^2\mathbf{E}-\mathbf{E}&=0 \\
    \frac{\omega^2-\omega_p^2+i\gamma\omega}{\beta^2}&=k^2.
\end{align}
The added solutions require an additional boundary condition that has been shown \cite{Moreau2013ImpactAntennas} to be satisfied by demanding that no currents leave the metal surface, and is expressed by
\begin{align}\label{RoughBoundaryCondition}
    \mathbf{J}_\perp = \mathbf{P}_{\perp \text{at each interface}} = 0,
\end{align}
where $\perp$ denotes the direction perpendicular to the surface. Combining Eq. \ref{RoughBoundaryCondition} with Eq. \ref{eq: AmpereEquation} yield a comprehensive solution, encompassing both the transverse and longitudinal contributions to the momentum dependent dispersion, while considering nonlocal effects.  

\section{Confinement and loss}
\begin{figure}[ht]
    \centering
    %\subfigure[]{\includegraphics[width=0.49\linewidth]{Au_chi.png}}
    \includegraphics[width=1\linewidth]{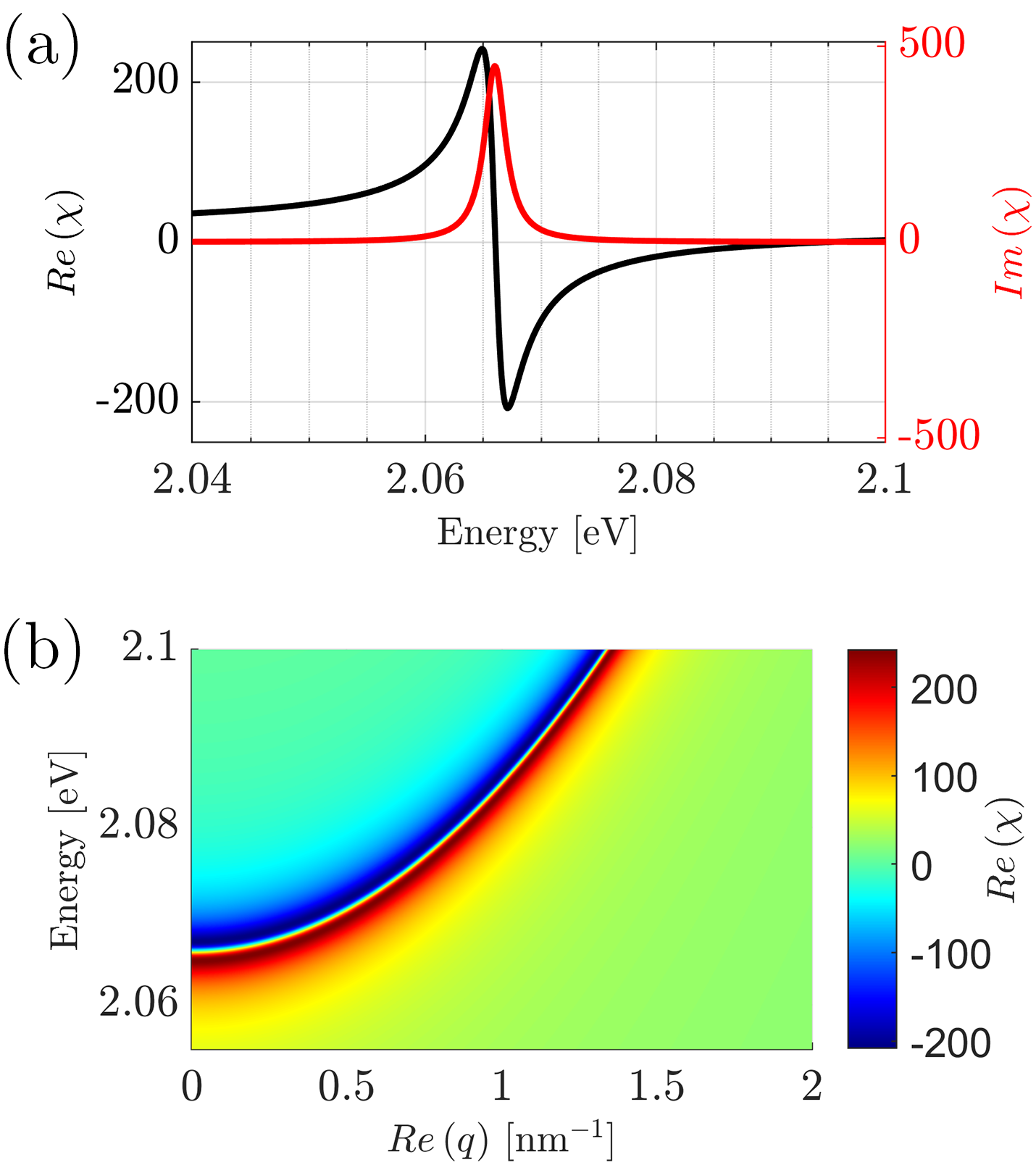}% Here is how to import EPS art
    \caption{Susceptibility $\chi_\textsubscript{TMD}$ of a hBN encapsulated WS\textsubscript{2} monolayer. (a) Real and imaginary parts of the susceptibility in the local model. (b) The real part of the susceptibility in color scale as a function of frequency and momentum in the nonlocal model.} 
    \label{fig:Figures dielectic function}
\end{figure}
The systems under investigation are illustrated in Fig. \ref{fig:Geometry}, depicting a monolayer WS\textsubscript{2} encapsulated by hBN on both surfaces (Fig. \ref{fig:Geometry}.a) and for comparison a similarly encapsulated 3 nm slab of Au (Fig. \ref{fig:Geometry}.b). Fig. \ref{fig:Figures dielectic function} presents the obtained susceptibility of the TMD for both the local (Fig. \ref{fig:Figures dielectic function}.a), and the nonlocal model (Fig. \ref{fig:Figures dielectic function}.b). For frequencies ranging from 2.065 to 2.1 eV, encapsulated WS\textsubscript{2} exhibits $Re\left(\chi_\textsubscript{TMD}\right)<1$ when cryogenically cooled, thus supporting the 2DEP mode in both local and nonlocal models \cite{Epstein2020HighlySemiconductors}. In the nonlocal model, $Re\left(\chi_\textsubscript{TMD}\right)<1$ is upheld for specific momenta in addition to specific spectral range. The momenta which uphold this conditions are considerably lower than those of the local model. To obtain the mode's dispersion, we numerically solve the resulting equation for TM surface waves \cite{Goncalves2016AnPlasmonics}:
\begin{align}\label{eq: TM Polariton dispersion equations}
    \frac{2\varepsilon_\textsubscript{hBN}}{\sqrt{q_{2DEP}^2-\varepsilon_\text{hBN}\frac{\omega^2}{c^2}}}+\chi_\textsubscript{TMD}d_0&=0,
\end{align}
where $\varepsilon_{hBN}\approx3.8756$ is the hBN's permittivity within these frequencies, $d_0$ is the WS\textsubscript{2} monolayer thickness, $q_{2DEP}$ is the 2DEP momentum component parallel to the TMD surface, i.e., in the direction of propagation, and $\chi_\text{WS\textsubscript{2}}$ is the TMD's susceptibility described by Eq. \ref{non_local_SC_susceptibility_single}. The parameter values for the hBN encapsulated WS\textsubscript{2} at 10K were taken from \cite{Epstein2020HighlySemiconductors} (see supplementary information). To describe the local solution, the "$\frac{\hbar^2k^2}{2M}$" term in Eq. \ref{non_local_SC_susceptibility_single} was omitted. 

To determine the dispersion of the encapsulated slab, we numerically compute the SRSPP dispersion supported by the given structure. This was done by solving Maxwell's equations with the revised Ampere's law (Eq. \ref{eq: AmpereEquation}) for a TM wave in a hBN/Au/hBN structure using boundary conditions as in Eq. \ref{RoughBoundaryCondition} and allowing both longitudinal and transverse components as in Eq. \ref{eq: metal_longitudinal_wavevector} to yield 
\begin{align}\label{eq: Au full dispersion}
    \begin{split} 
            0 &= \left[C_\text{Au}C_\text{hBN}\left(B+A_\text{hBN}+A_\text{Au}\right)\right]^2+\\
            &-\left[C_\text{hBN}\left(A_\text{Au}-A_\text{hBN}-B\right)\right]^2+\\
            &-\left[C_\text{Au}\left(A_\text{hBN}+A_\text{Au}-B\right)\right]^2+\\
            &-8C_\text{Au}C_\text{hBN}A_\text{Au}B+\left(A_\text{Au}-A_\text{hBN}+B\right)^2
    \end{split},
\end{align}
 where
 \begin{align}
     \begin{matrix}
         A_\text{hBN}=\frac{\sqrt{q^2-\varepsilon_\text{hBN}k_0^2}}{\varepsilon_\text{hBN}} & A_\text{Au}=\frac{\sqrt{q^2-\varepsilon_\text{Au}k_0^2}}{\varepsilon_\text{Au}} \\ B=\frac{q^2}{\sqrt{q^2+\frac{1}{\alpha}}}\left(\frac{1}{\varepsilon_\text{Au}}-1\right) &  \alpha=\frac{\beta^2}{\omega_p^2}\left(1-\frac{1}{\varepsilon_\text{Au}}\right) \\ C_\text{Au}=e^{d\sqrt{\varepsilon_\text{Au}k_0^2-q^2}} & C_\text{hBN}=e^{d\sqrt{\varepsilon_\text{hBN}k_0^2-q^2}} 
     \end{matrix},
 \end{align}
where $k_0=\omega/c$ is the momentum in vacuum and $d$ is the Au slab thickness. The values for the permittivity of Au, $\varepsilon_{Au}\left(\omega\right)$ were taken from Johnson and Christy \cite{JohnsonChristy}. The values for $\omega_p$ and $\gamma$ were obtained by fitting the experimentally measured values \cite{JohnsonChristy} to the Drude-like model $\varepsilon=1-\frac{\omega_p^2}{\omega^2+i\gamma\omega}$ as in \cite{Moreau2013ImpactAntennas}. For the case without nonlocal considerations the solution is similar for $\alpha=0$. The modes and dispersion relations $\omega(k)$ of both polaritons are achieved by numerically solving the dispersion relations given by Eq. \ref{eq: TM Polariton dispersion equations} and \ref{eq: Au full dispersion}.
\begin{figure}[ht]
    \centering
    \includegraphics[width=1\linewidth]{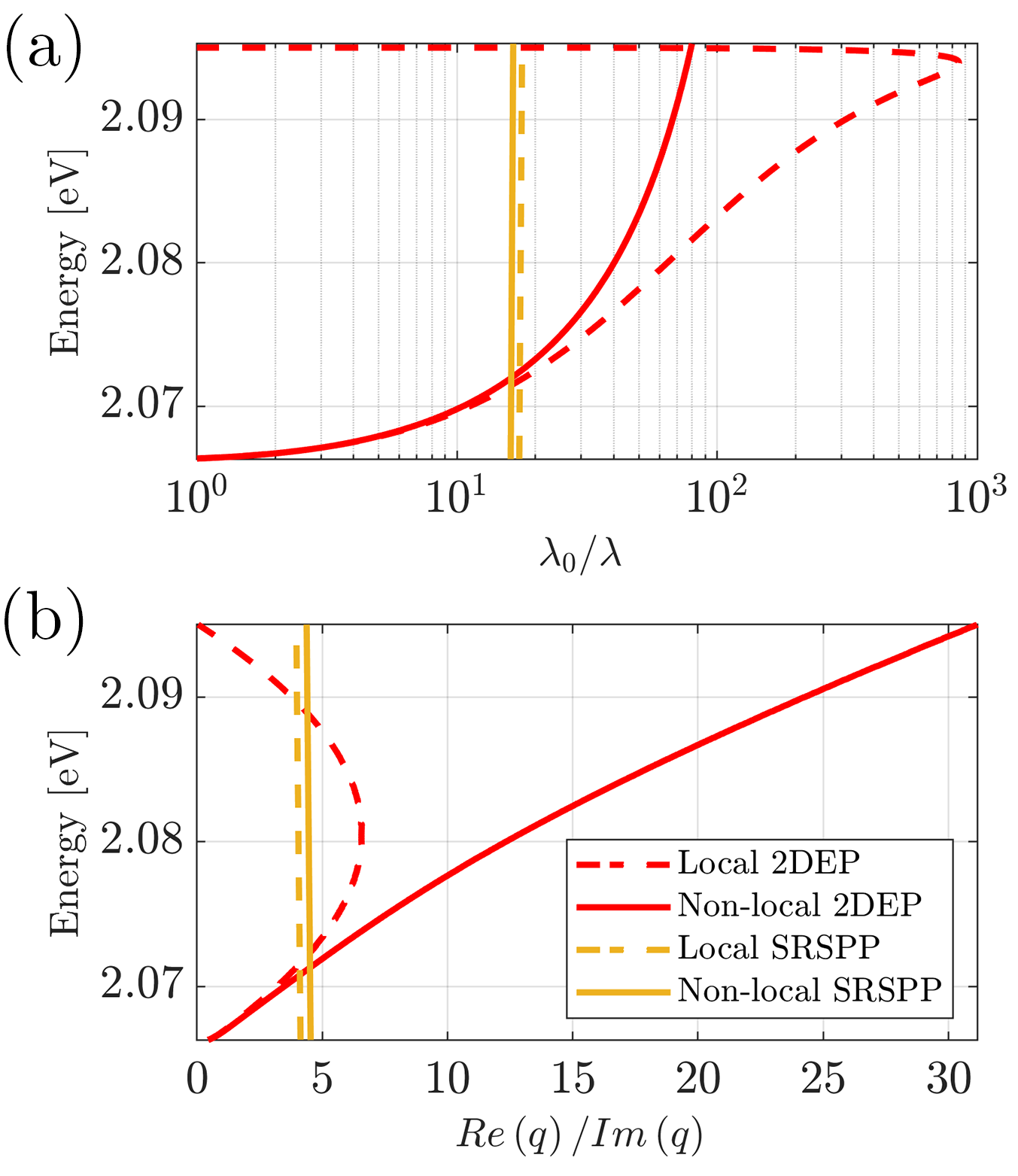}% Here is how to import EPS art
    %\vspace{0.00mm}    \subfigure{\includegraphics[width=1\linewidth]{Legend.png}}
    \caption{Confinement and loss 2DEPs and SRSPP calculated for both local (dashed lines) and nonlocal models (solid lines). (a) Confinement factor. (b) Propagation losses figure of merit}  
    \label{fig:Figures Dispersion and FOM}
\end{figure}

Fig. \ref{fig:Figures Dispersion and FOM}.a presents the modes’ confinement factor,  $\lambda _0/\lambda$, as a function of energy, where $\lambda_0$ is the wavelength in vacuum, and $\lambda$ is the modes' wavelength. This quantity is an indication of how strong the polaritonic modes are confined since higher momenta modes translates to shorter wavelengths and thus higher confinement factors. This is done for both the local (dashed line) and nonlocal (solid line) models for the 2DEP (red) and SRSPP (gold). It can be seen that for most frequencies shown, the 2DEP exhibits a higher confinement factor compared to that of the SRSPP, by up to an order of magnitude. Fig. \ref{fig:Figures Dispersion and FOM}.b shows the ratio between the real and imaginary parts of the momentum, which is the figure of merit representing the number of wavelengths a mode propagates before decaying \cite{Woessner2014}. Fig. \ref{fig:Figures Dispersion and FOM}.b shows the great advantage of the 2DEP by demonstrating that even at high momenta they have six times less losses compared to the SRSPP at the same frequency. 

%\pagebreak
\section{Polaritonic coupling to the Valley degree of freedom}
\begin{figure}[ht]
    \centering
    \includegraphics[width=1\linewidth]{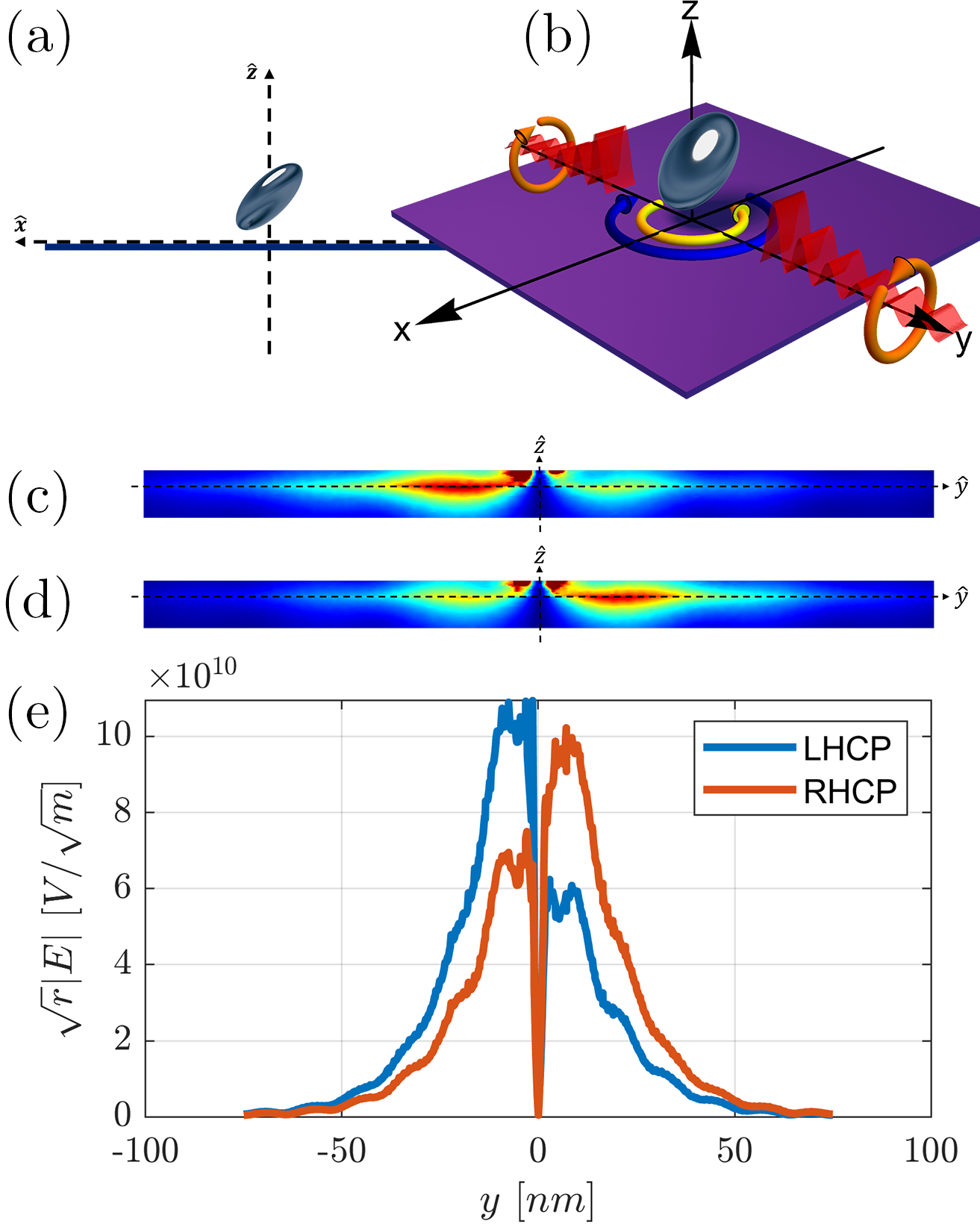}% Here is how to import EPS art

    \caption{Valley coupled, highly directional 2DEPs on a monolayer WS\textsubscript{2}. Illustrated system schematic depicting the nano particle situated above the axis origin, tilted at an angle. Circularly polarized excitonic dipole states in the in-plane direction are presented by blue/yellow curves. Red ribbons illustrate left/right propagating 2DEPs with left/right circular polarization in the out-of-plane direction in orange. Side (a) and perspective (b) views. Cross section in the $y-z$ of the magnitudes of the electric field for left (c) and right (d) handed polarizations. (e) Same cross section view of the magnitude of the electric field as a function of distance in meters $y$ for $z=0$, from the nanoparticle's center position.}
    
    \label{fig:Valley Simulations}
\end{figure}
Next, we explore the possible effect of the valley degree of freedom on the polaritonic modes. In monolayer TMDs, the combination of lattice inversion asymmetry, time reversal symmetry, and strong spin-orbit coupling leads to different optical selection rules for excitons residing at the different \(K\) and \(K'\) valleys \cite{Xiao2012CoupledDichalcogenides}. Thus, excitons in different valleys are optically accessible by the associated left or right circular polarizations of the incident or emitted light \cite{Mak2012ControlHelicity,Sallen2012RobustExcitation}. The excitons’ polarization orientation is in-plane with respect to the 2D material surface plane (yellow/blow curves in the x-y plane in Fig. \ref{fig:Valley Simulations}.b). Since the excitons are coupled to the valley degree of freedom, and the 2DEPs are coupled to the excitons, we expect that the valley polarization will also affect the 2DEPs. In contrast, noble metals such as Au and Ag, which support surface plasmons, do not posses a valley degree of freedom property. However, a general attribute of any surface polariton is that it is polarized in the out-of-plane direction with respect to the surface plane, i.e., perpendicular to the TMD exciton’s polarization plane (orange curves in Fig. \ref{fig:Valley Simulations}.b). These polarizations are correlated with asymmetric directional propagation, such that opposite circular polarizations (left/right) are directly tied with opposite directions of propagation, as has been previously demonstrated for plasmonic modes \cite{Rodriguez-Fortuno2013Near-fieldModes, OConnor2014SpinorbitNanostructures}. To observe the valley dependency of the 2DEPs, one needs to couple the exciton’s and the 2DEP’s electric fields from two different perpendicular planes (Fig. \ref{fig:Valley Simulations}.a and \ref{fig:Valley Simulations}.b). While coupling between the excitons' valley property to directional propagation has been previously observed by utilizing several asymmetric geometries \cite{Sun2018TuningSystems,Chervy2018RoomPlasmons,Gong2018NanoscaleCoupling}, here we show that 2DEPs are also coupled to the valley degree of freedom. This, in turn, should be reflected by an asymmetric directional propagation of the 2DEP to either left or right side, according to either left or right circular polarization (red curves in Fig. \ref{fig:Valley Simulations}), as have been previously demonstrated for hyperbolic exciton polaritons in few-layer TMDs \cite{Sternbach2021,Eini2022Valley-polarizedFrequencies}.

We achieve the coupling of the excitonic and polaritonic modes by means of a tilted asymmetrically aligned ellipsoid nanoparticle excitation with regards to the surface of the TMD (Fig. \ref{fig:Valley Simulations}.a,b). In order to couple to such modes accounting for the momentum-mismatch, and to accurately predict and calculate its properties, nonlocality has to be considered, as in Eq. \ref{eq: non_local_SC_susceptibility}. This, will manifest in the dispersion of the excited mode. To explore this possibility, we have used COMSOL MULTIPHYSICS to simulate the interaction of a circularly handed dipole, representing excitonic emission from a specific valley in the $x-y$ plane, with the ellipsoid particle tilted around $\hat{y}$ in the $x-z$ plane. The nanoparticle's permittivity was taken as $\varepsilon=-18-4j$ and for $E=2.08 eV$ relating to a specific mode a the 2DEP. Fig. \ref{fig:Valley Simulations}.c,d show the absolute value of the electric field in a $y-z$ cross section for a left and right handed circular polarization excitations respectively, corresponding to different valleys. The propagation direction of the launched 2DEPs'  electric field is evident to be in the $\mp\hat{y}$ directions and in correlation with the left/right polarization handedness. In Fig. \ref{fig:Valley Simulations}.e we show the magnitude of the electric field along the $z=0$ line, multiplied by the square root of the distance to account for the energy decay in a 2D surface wave. It can be clearly seen that a preferable propagation in the left or right direction from the center of the nanoparticle is obtained from the left or right circularly polarized dipole excitation. Thus, we find that unlike SPPs, 2DEPs are coupled and affected by the valley degree of freedom in TMDs. 

%\pagebreak
\section{Conclusion}
In this work, we have performed a comprehensive study and comparison of the polaritonic properties of 2DEPs and SRSPPs, including the appropriate nonlocal corrections to each system. Our findings clearly show the advantages of 2DEPs over those of SRSPPs, offering both high confinement and reduced losses, making them promising candidates for nanophotonic applications in the VIS/NIR spectral range. In addition, we have demonstrated that unlike SPPs, 2DEPs are coupled to the valley degree of freedom, harnessing this unique property. The accurate nonlocal treatment of these two systems ensures a comprehensive understanding of the phenomena at hand, with the precise predictions of their actual polaritonic properties.
%\pagebreak
\section{ACKNOWLEDGMENTS}
This work was supported by the European Union (ERC, TOP-BLG, project 101078192). The authors would like to thank Dr. Yarden Mazor and Tomer Eini for fruitful discussions and assistance with COMSOL simulations.

%\pagebreak
%\clearpage
\bibliography{references}
\end{document}